\documentclass[sigconf]{acmart}
\AtBeginDocument{%
  \providecommand\BibTeX{{%
    \normalfont B\kern-0.5em{\scshape i\kern-0.25em b}\kern-0.8em\TeX}}}

\usepackage{subfig}
\usepackage{textgreek}
\usepackage{multirow}
\usepackage{xcolor}
\usepackage{listings}
\usepackage[ruled,linesnumbered]{algorithm2e}

\copyrightyear{2022}
\acmYear{2022}
\setcopyright{acmcopyright}\acmConference[ICSE '22]{44th International Conference on Software Engineering}{May 21--29, 2022}{Pittsburgh, PA, USA}
\acmBooktitle{44th International Conference on Software Engineering (ICSE '22), May 21--29, 2022, Pittsburgh, PA, USA}
\acmPrice{15.00}
\acmDOI{10.1145/3510003.3510096}
\acmISBN{978-1-4503-9221-1/22/05}

\begin{document}

\title{SPT-Code: Sequence-to-Sequence Pre-Training for Learning Source Code Representations}


\author{Changan Niu}
\affiliation{
  \institution{State Key Laboratory for Novel Software Technology\\Nanjing University}
  \streetaddress{22 Hankou Road}
  \city{Nanjing}
  \country{China}}
\email{nougatca@qq.com}

\author{Chuanyi Li}
\affiliation{
  \institution{State Key Laboratory for Novel Software Technology\\Nanjing University}
  \streetaddress{22 Hankou Road}
  \city{Nanjing}
  \country{China}}
\email{lcy@nju.edu.cn}

\author{Vincent Ng}
\affiliation{
  \institution{Human Language Technology Research Institute\\University of Texas at Dallas}
  \streetaddress{Richardson, TX 75083-0688}
  \city{Richardson}
  \state{Texas}
  \country{USA}}
\email{vince@hlt.utdallas.edu}

\author{Jidong Ge}
\affiliation{
  \institution{State Key Laboratory for Novel Software Technology\\Nanjing University}
  \streetaddress{22 Hankou Road}
  \city{Nanjing}
  \country{China}}
\email{gjd@nju.edu.cn}

\author{Liguo Huang}
\affiliation{
  \institution{Dept. of Computer Science\\Southern Methodist University}
  \streetaddress{Dallas, TX 75275-0122}
  \city{Dallas}
  \state{Texas}
  \country{USA}}
\email{lghuang@lyle.smu.edu}

\author{Bin Luo}
\affiliation{
  \institution{State Key Laboratory for Novel Software Technology\\Nanjing University}
  \streetaddress{22 Hankou Road}
  \city{Nanjing}
  \country{China}}
\email{luobin@nju.edu.cn}


\begin{abstract}
    Recent years have seen the successful application of large pre-trained models to code representation learning, resulting in substantial improvements on many code-related downstream tasks. But there are issues surrounding their application to SE tasks. First, the majority of the pre-trained models focus on pre-training only the encoder of the Transformer. For generation tasks that are addressed using models with the encoder-decoder architecture, however, there is no reason why the decoder should be left out during pre-training. Second, many existing pre-trained models, including state-of-the-art models such as T5-learning, simply reuse the pre-training tasks designed for natural languages. 
    Moreover, to learn the natural language description of source code needed eventually for code-related tasks such as code summarization, existing pre-training tasks require a bilingual corpus composed of source code and the associated natural language description, which severely limits the amount of data for pre-training. To this end, we propose SPT-Code, a sequence-to-sequence pre-trained model for source code. In order to pre-train SPT-Code in a sequence-to-sequence manner and address the aforementioned weaknesses associated with existing pre-training tasks, we introduce three pre-training tasks that are specifically designed to enable SPT-Code to learn knowledge of source code, the corresponding code structure, as well as a natural language description of the code without relying on any bilingual corpus, and eventually exploit these three sources of information when it is applied to downstream tasks. Experimental results demonstrate that SPT-Code achieves state-of-the-art performance on five code-related downstream tasks after fine-tuning.
\end{abstract}




\maketitle

\section{Introduction}
\label{section:introduction}

Pre-Training has revolutionized the way computational models are trained in the natural language processing (NLP) community~\cite{devlin2019bert,radford2019language,peters2018deep,clark2019electra}. For a long time, supervised learning has been the most successful natural language learning paradigm. The pioneers of the pre-training idea challenged this view by showing that a vast amount of general knowledge about language, including both linguistic and commonsense knowledge, can be acquired by (pre-)training a model in a {\em task-agnostic} manner using {\em self-supervised} learning tasks. Self-supervised learning tasks are NLP tasks for which the label associated with a training instance can be derived automatically from the text itself. Consider, for instance, one of the most well-known self-supervised learning tasks, Masked Language Modeling (MLM)~\cite{devlin2019bert}. Given a sequence of word tokens in which a certain percentage of tokens is {\em masked} randomly, the goal of MLM is to predict the masked tokens. As can be easily imagined, a model for MLM can therefore be trained on instances where each one is composed of a partially masked sequence of word tokens and the associated ``class'' value is the masked tokens themselves. Because no human annotation is needed, a model can be pre-trained on a very large amount of labeled data can be automatically generated, thereby acquiring a potentially vast amount of knowledge about language. A pre-trained model can then be optimized for a specific task by fine-tuning its parameters using task-specific labeled data in the standard supervised fashion.

Inspired by the successes of pre-trained models in NLP, a number of pre-trained models for source code have been proposed and applied to a variety of SE tasks including code summarization and code completion, with notable successes~\cite{buratti2020exploring,karampatsis2020scelmo,kanade2020learning,svyatkovskiy2020intellicode,feng2020codebert,guo2020graphcodebert,liu2020multi,mastropaolo2021studying,jiang2021treebert}. Despite these promising results, there are issues surrounding the application of pre-trained models for source code to SE tasks. 

First, the majority of these pre-trained models focus on pre-training only the encoder of the Transformer~\cite{kanade2020learning,buratti2020exploring,feng2020codebert,guo2020graphcodebert}. This is not ideal, however. For instance, for generation tasks that are addressed using models with the encoder-decoder architecture, there is no reason why the decoder should be left out in the pre-training process. Second, these models have largely assumed as inputs the source code~\cite{kanade2020learning,karampatsis2020scelmo,svyatkovskiy2020intellicode,buratti2020exploring,liu2020multi} and the associated natural language description~\cite{feng2020codebert,mastropaolo2021studying}. In particular, code structure, which is also crucial to understanding source code, is largely missing from these models. The reason why code structure is left out is that SE researchers have for the most part simply reused the pre-training tasks designed for natural languages when pre-training models (by viewing source code as natural language), but none of these tasks are concerned with learning the structure of a
natural language. Third, these pre-training tasks assume the availability of a bilingual corpus, where each method/function (henceforth collectively referred to as method) is ``labeled'' with the corresponding docstring, when pre-training a model on source code and the associated natural language description~\cite{feng2020codebert,guo2020graphcodebert}. However, such a bilingual corpus tends to be small in size compared to a monolingual (i.e., source code only) corpus, thus severely limiting the amount of data a model can be pre-trained on. 
In general, we believe the reliance on a bilingual corpus would hinder the development of powerful pre-trained models for source code in the long run, as a key strength of the pre-trained models developed in the NLP community stems from their ability to be trained using self-supervised tasks for which \textit{very large} amounts of labeled data can be automatically generated.

Several attempts have been made to address the three problems mentioned above to different extents. To address the first problem, T5-learning~\cite{mastropaolo2021studying} and TreeBERT~\cite{jiang2021treebert} are proposed, which are  sequence-to-sequence (i.e., seq2seq) pre-training models with the encoder-decoder architecture and enables both the encoder and the decoder to be jointly trained. To address the second problem, GraphCodeBERT~\cite{guo2020graphcodebert} employs two pre-training tasks specifically designed to acquire structural information. One task involves predicting the edges in the data flow while the other involves predicting the alignment between the nodes in the data flow and the code sequence, respectively. However, as the authors also pointed out, while the data flow captures information that is largely semantic in nature, it does not capture syntactic information (e.g., the syntactic structure encoded in an Abstract Syntax Trees, i.e., AST), which is arguably the most important type structural information about source code that is commonly exploited by SE researchers. 
TreeBERT~\cite{jiang2021treebert} employs the set of constituent paths of ASTs as the input of its encoder. However, it only inputs code sequences at the decoder side during pre-training, trying to make the encoder learn lexical and semantic information (both of which can be easily obtained from the code sequences) from the AST (which contains mainly syntactic information) during the pre-training phase. But it is uncertain whether TreeBERT can extract the complete lexical and semantic information from the AST only by relying on pre-training, thus eliminating the need to input code tokens when fine-tuning.
To address the third problem, T5-learning treats code and natural language as two types of independent data instances. While this allows T5-learning to learn from a monolingual rather than bilingual corpus, the connection between a piece of code and the associated natural language description is no longer present in the corpus. Hence, it is no longer clear whether T5-learning can still learn to produce a natural language description of a piece of code. To our knowledge, there has been no attempt to address all three issues in a single model.

In light of the above discussion, we propose SPT-Code, a new pre-trained model for source code. Motivated by T5-learning, SPT-Code is a seq2seq pre-training model, enabling both the encoder and the decoder of Transformer to be jointly pre-trained. Each data instance for SPT-Code is composed of three types of information derived from a method, namely the code sequence, its AST, and the associated natural language description. Note that the incorporation of ASTs allows SPT-Code to exploit structural, specifically syntactic, information. In addition, to obviate the need to learn natural language descriptions from a bilingual corpus, we will simply use the name of the method and the names of the methods that are invoked in this method as a (very succinct) natural language description of the given source code. 

We design three code-specific pre-training tasks for SPT-Code, each of which allows SPT-Code to acquire exactly one of the three types of information that comprise a data instance. The first task is a version the well-known Masked Sequence to Sequence (MASS)~\cite{song2019mass} pre-training task for natural language that we adapt to source code. Specifically, our modified MASS task seeks to acquire knowledge about source code via masking a random fragment of the code tokens. The second task, Code-AST prediction (CAP), is designed to enable the model to gain knowledge of the syntactic structure of a code fragment by predicting whether the given AST is the correct AST for the given piece of code. The final task, Method Name Generation (MNG), is a novel task that involves generating the subtokens of the method name, which we take to be an (extremely concise) natural language description of the method.

After SPT-Code is pre-trained on the CodeSearchNet dataset~\cite{husain2020codesearchnet}, we fine-tune and evaluate it on five downstream tasks, including code summarization, code completion, bug fixing, code translation, and code search. Experimental results show that SPT-Code achieves state-of-the-art results under virtually all circumstances. 

In sum, we make the following contributions:
\begin{enumerate}
    \item Propose SPT-Code, a seq2seq pre-trained model for source code that is built with the encoder-decoder architecture and is applicable to both classification and generation tasks.
    \item Extend the input representation of pre-trained models for source code with a simplified and linearized version of ASTs. To our knowledge, we are the first to use both natural language and AST as inputs in pre-training for source code.
    \item Design special input representations and three code-specific seq2seq-based pre-training tasks enabling SPT-Code to be pre-trained without relying on any bilingual corpus or labeled data. 
    \item Pre-train SPT-Code on a large unlabeled monolingual (i.e., source code only) dataset across six programming languages, then fine-tune and evaluate it on five downstream code-related tasks, achieving state-of-the-art results on all tasks.
\end{enumerate}

\section{Related Work}
\label{section:related}

\subsection{Pre-Training Models in NLP and SE}

Table~\ref{table:pre_train_models}
\begin{table*}
    \centering
    \caption{Pre-training models in NLP and for source code.}
    \label{table:pre_train_models}
    \resizebox{\textwidth}{!}{
    \begin{tabular}{llcccc|cccccccc|ccc}
    \toprule
        \multicolumn{1}{l}{\multirow{2}{*}{Domain}} &
        \multicolumn{1}{l}{\multirow{2}{*}{Models}} &
        \multicolumn{4}{c|}{Modules} &
        \multicolumn{8}{c|}{Objectives} &
        \multicolumn{3}{c}{Inputs} \\
    \cline{3-17}
        \multicolumn{1}{c}{} &
        \multicolumn{1}{c}{} &
        LSTM &
        Encoder &
        Decoder &
        Encoder-Decoder &
        Forward LM &
        Backward LM &
        Masked LM &
        NSP &
        Permutation LM &
        RTD &
        - &
        - &
        NL &
        Code &
        Structure \\
    \midrule
        \multirow{8}{*}{NLP}  & ELMo (2018)     & \checkmark &            &            &            & \checkmark & \checkmark &            &            &            &            & & & \checkmark & & \\ \cline{2-17}
                              & BERT (2019)     &            & \checkmark &            &            &            &            & \checkmark & \checkmark &            &            & & & \checkmark & & \\ \cline{2-17}
                              & XLNet (2019)    &            & \checkmark &            &            &            &            &            &            & \checkmark &            & & & \checkmark & & \\ \cline{2-17}
                              & RoBERTa (2019)  &            & \checkmark &            &            &            &            & \checkmark &            &            &            & & & \checkmark & & \\ \cline{2-17}
                              & ELECTRA (2019)  &            & \checkmark &            &            &            &            & \checkmark &            &            & \checkmark & & & \checkmark & & \\ \cline{2-17}
                              & GPT-2 (2019)    &            &            & \checkmark &            & \checkmark &            &            &            &            &            & & & \checkmark & & \\ \cline{2-17}
                              & T5 (2020)       &            &            &            & \checkmark &            &            & \checkmark &            &            &            & & & \checkmark & & \\ \cline{2-17}
                              & BART (2020)     &            &            &            & \checkmark &            &            & \checkmark &            &            &            & & & \checkmark & & \\
    \hline
        \multirow{7}{*}{Code} & SCEMLo (2020)        & \checkmark &            &            &            & \checkmark & \checkmark &            &            &            &            &      &     &            & \checkmark &           \\ \cline{2-17}
                              & CuBERT (2019)        &            & \checkmark &            &            &            &            & \checkmark & \checkmark &            &            &      &     &            & \checkmark &           \\ \cline{2-17}
                              & C-BERT (2020)        &            & \checkmark &            &            &            &            & \checkmark & \checkmark &            &            &      &     &            & \checkmark &           \\ \cline{2-17}
                              & IntelliCode (2020)   &            &            & \checkmark &            &            & \checkmark &            &            &            &            &      &     &            & \checkmark &           \\ \cline{2-17}
                              & CodeBERT (2020)      &            & \checkmark &            &            &            &            & \checkmark &            &            & \checkmark &      &     & \checkmark & \checkmark &           \\ \cline{2-17}
                              & GraphCodeBERT (2020) &            & \checkmark &            &            &            &            & \checkmark &            &            &            & EP   & NA  & \checkmark & \checkmark & Data Flow \\ \cline{2-17}
                              & CugLM (2020)         &            & \checkmark &            &            & \checkmark &            & \checkmark & \checkmark &            &            &      &     &            & \checkmark &           \\ \cline{2-17}
                              & T5-learning (2021)   &            &            &            & \checkmark &            &            & \checkmark &            &            &            &      &     & \checkmark & \checkmark &           \\ \cline{2-17}
                              & TreeBERT (2021)      &            &            &            & \checkmark &            &            &            &            &            &            & TMLM & NOP &            & \checkmark & AST       \\
    \bottomrule
    \end{tabular}
    }
\end{table*}
presents an overview of the most prominent pre-trained models in NLP and for code. Each model is characterized along four dimensions: (1) Modules: what is being pre-trained (e.g., the encoder, the decoder, or both); (2) Objectives: the pre-training objectives\footnote{Forward LM~\cite{peters2018deep,radford2019language,karampatsis2020scelmo} aims to predict the next word given the preceding words in a sentence. Backward LM~\cite{peters2018deep,karampatsis2020scelmo} aims to predict the previous word given the words that appear after it in a sentence. Masked LM is the masked language modeling task described in Section~\ref{section:introduction}. NSP~\cite{devlin2019bert} aims to predict whether the second sentence in a sentence pair should immediately follow the first sentence in the pair. Permutation LM~\cite{yang2019xlnet} aims to predict a word using a set of context words randomly selected via the attention mask mechanism. RTD aims to predict which token in the input has been replaced. EP~\cite{guo2020graphcodebert} involves masking 20\% of the edges in a data flow and aims to predict the masked edges. NA~\cite{guo2020graphcodebert} involves masking a certain portion of edges of connecting data flow nodes and code tokens and aims to predict the masked edges. TMLM~\cite{jiang2021treebert} masks paths in the AST input on the encoder side and tokens in the code sequence input on the decoder side, then predicts the token of the masked code. NOP~\cite{jiang2021treebert} is to exchange the positions of some nodes in the path, and distinguish whether the order of nodes in the AST is correct or not.} and (3) Inputs: what information the model assumes as input (e.g., natural language, code and structural information of the code).

We can make a few observations from Table~\ref{table:pre_train_models}. First, while the majority of work has focused on pre-training the encoder, the newest models (T5~\cite{raffel2020exploring}, BART~\cite{lewis2020bart}, and T5-learning~\cite{mastropaolo2021studying}, which is modeled after T5) are all seq2seq pre-training models that allow the encoder and the decoder to be jointly trained. Second, except for GraphCodeBERT~\cite{guo2020graphcodebert} and TreeBERT~\cite{jiang2021treebert}, all pre-trained models in SE reuse training objectives designed for natural languages, with MLM being the most popular pre-training task. This indicates that the selection of pre-training tasks, which dictates what knowledge will be acquired and exploited by a model, is an area of research that is under-investigated in SE. Finally, while earlier pre-training models in SE assume only source code as input, the later ones all use both code and language.

In our experiments, we will use as our baselines the most recently developed (and also the state-of-the-art) pre-trained models for source code, namely CodeBERT~\cite{feng2020codebert}, GraphCodeBERT~\cite{guo2020graphcodebert}, CugLM~\cite{liu2020multi}, T5-learning~\cite{mastropaolo2021studying} and TreeBERT~\cite{jiang2021treebert}.

\subsection{Structural Information of Source Code}
\label{section:related_structure}

Structural information is very important for understanding source code. ASTs are widely used in code related tasks for representing structure information of code (e.g., ~\cite{zhang2020retrieval,wang2021code,svyatkovskiy2019pythia,hu2018deep,leclair2019neural}), which contains abundant syntactic structure information that cannot be expressed by code sequences. An AST should be flattened with linearization methods, e.g., pre-order traversal~\cite{zhang2020retrieval,wang2021code}, in-order traversal~\cite{svyatkovskiy2019pythia}, and Structure-based Traversal (SBT)~\cite{hu2018deep}, before being fed to an encoder. code2vec~\cite{alon2019code2vec}, code2seq~\cite{alon2018code2seq}, and SLM~\cite{alon2020structural} use a method that linearizes an AST as a series of ``path-contexts'' representing two terminal nodes and the path between them. Jiang et al.~\cite{jiang2021treebert} represent ASTs as the set of paths and then introduce node position embedding to obtain the position of the node in the AST. Besides, neural networks that take trees as input (e.g., tree-LSTMs~\cite{wan2018improving,wan2019multi,lin2021improving}, RvNNs~\cite{zhang2019novel} and GNNs/GCNs~\cite{yang2021multi,leclair2020improved,chen2021holistic}) utilize an AST directly instead of flattening it. 

There are also approaches taking data flow and control flow extracted from code as structural information, e.g., ~\cite{guo2020graphcodebert} and~\cite{huo2020control}. However, these flows do not contain structural information as rich as ASTs~\cite{guo2020graphcodebert}. Their advantages over ASTs are that they have a lower demand on hardware and need less training time under the same conditions~\cite{guo2020graphcodebert}. 

Taken together, we choose to use ASTs to represent the structural information in SPT-Code.

\section{SPT-Code}

In this section we first introduce the architecture of SPT-Code (Section~\ref{section:architechture}). We then describe the model input (Section~\ref{section:model_input}) and the pre-training tasks (Section~\ref{section:pre_training_tasks}), which are the key innovations of this paper. Finally, we illustrates how to fine-tune SPT-Code when it is applied to downstream tasks (Section~\ref{section:fine-tuning}).

\subsection{Model Architecture}
\label{section:architechture}

Architecturally, SPT-Code is essentially a multi-layer Transformer~\cite{vaswani2017attention}, which is widely used by pre-training models such as BART~\cite{lewis2020bart} and T5~\cite{raffel2020exploring}. As far as the parameter setting of the encoder and decoder is concerned, SPT-Code follows CodeBERT and GraphCodeBERT: (1) the number of layers (i.e., Transformer blocks) $L=12$, (2) the size of model $d_{model}=768$, (3) the dimension of feed forward $d_{ff}=3072$, (4) the number of self-attention heads $h=12$, and (4) the dropout rate $p_{dropout}=0.1$. The total number of parameters is 262M.

To pre-train both classification and generation tasks with an encoder-decoder structure, we adopt the strategy used in BART. In particular, note that the encoder and decoder will continue to learn jointly and collaboratively when pre-trained on classification tasks.

\begin{figure}[htbp]
  \centering
  \subfloat[When SPT-Code is used for classification, the inputs of the encoder and decoder are identical, and the output of the decoder at the last time step is used as the label for the classification.]{\includegraphics[width=0.4\textwidth]{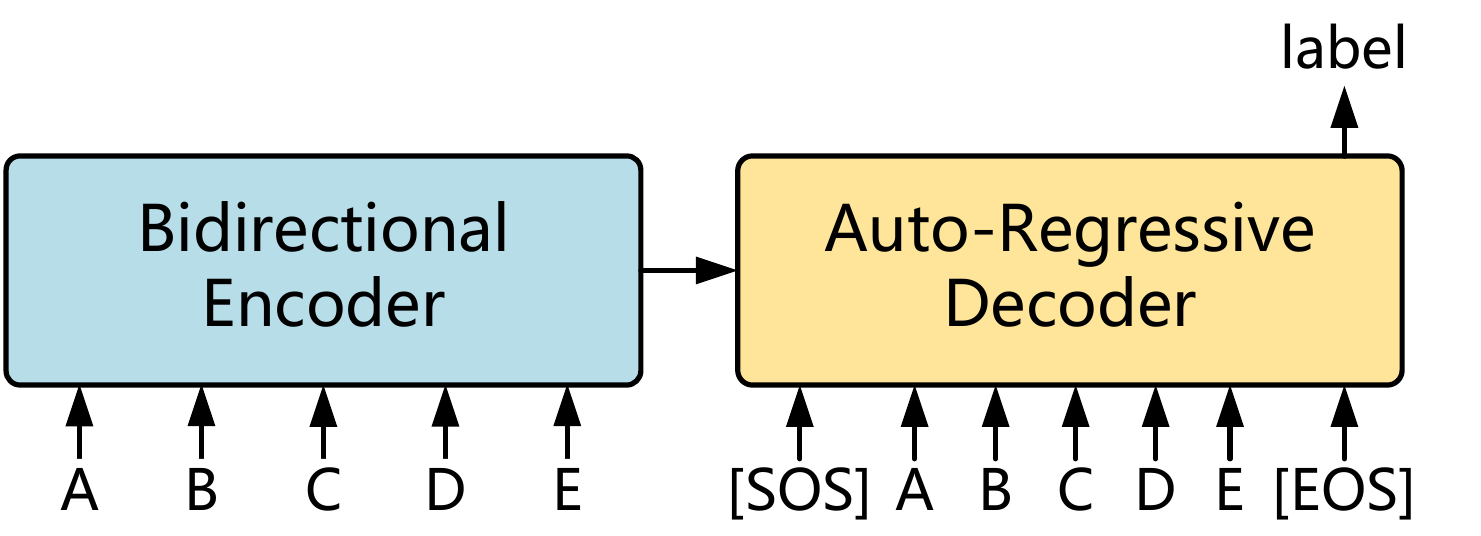}\label{figure:model_cls}}\\
  \subfloat[When SPT-Code is used for generation, the whole procedure is the same as the regular sequence-to-sequence model.]{\includegraphics[width=0.4\textwidth]{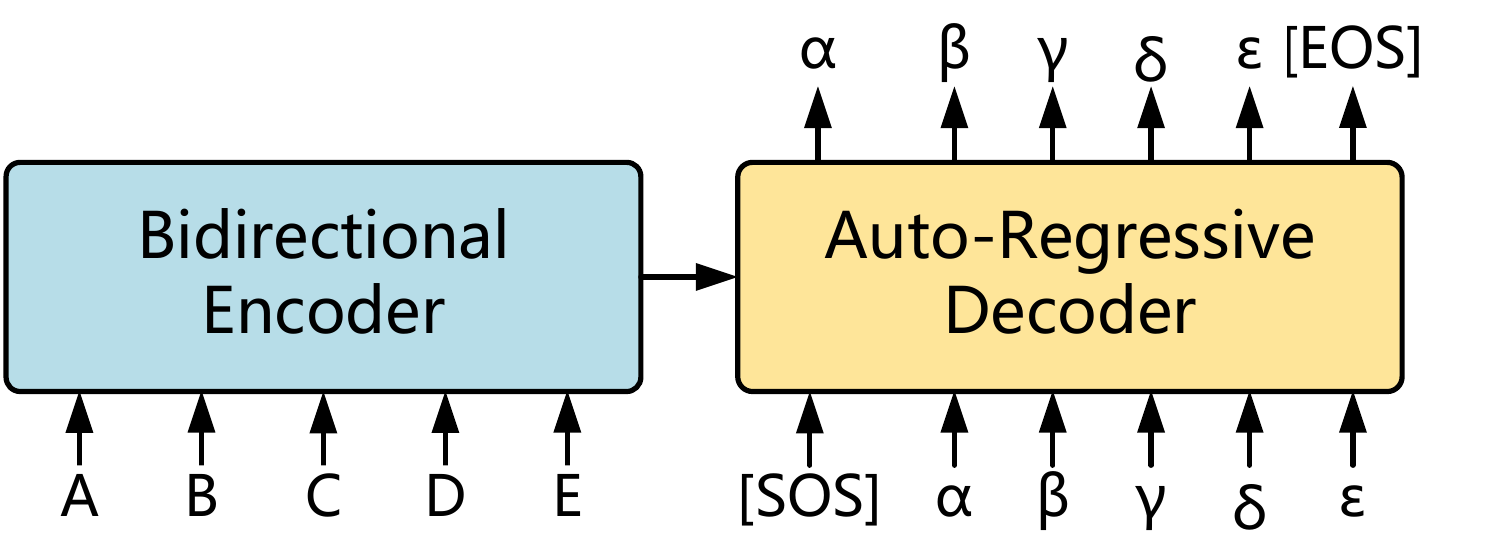}\label{figure:model_gen}}
  \caption{SPT-Code for classification and generation.}
  \Description{Input and output of SPT-Code for classification and generation.}
  \label{figure:model_cls_gen}
\end{figure}

Figure~\ref{figure:model_cls_gen} shows how classification and generation tasks are pre-trained in SPT-Code. Specifically, 
to use SPT-Code for classification tasks, the input of the decoder is the same as encoder, except that a special symbol ``[SOS]'' (indicating the start of the sequence) is added to the front, and another special symbol ``[EOS]'' (used here as a placeholder for the classification position) is added to the end. The output of the corresponding position of the ``[EOS]'' is then used for classification. For generation tasks, Figure~\ref{figure:model_gen} demonstrates the input and output of the model when translating ``A B C D E'' to ``\textalpha~\textbeta~\textgamma~\textdelta~\textepsilon''. The special symbol ``[EOS]'' denotes the end of the sequence, and the process of generation stops when the model outputs this symbol.

\subsection{Model Input}
\label{section:model_input}

\begin{figure}[htbp]
  \centering
  \includegraphics[width=0.47\textwidth]{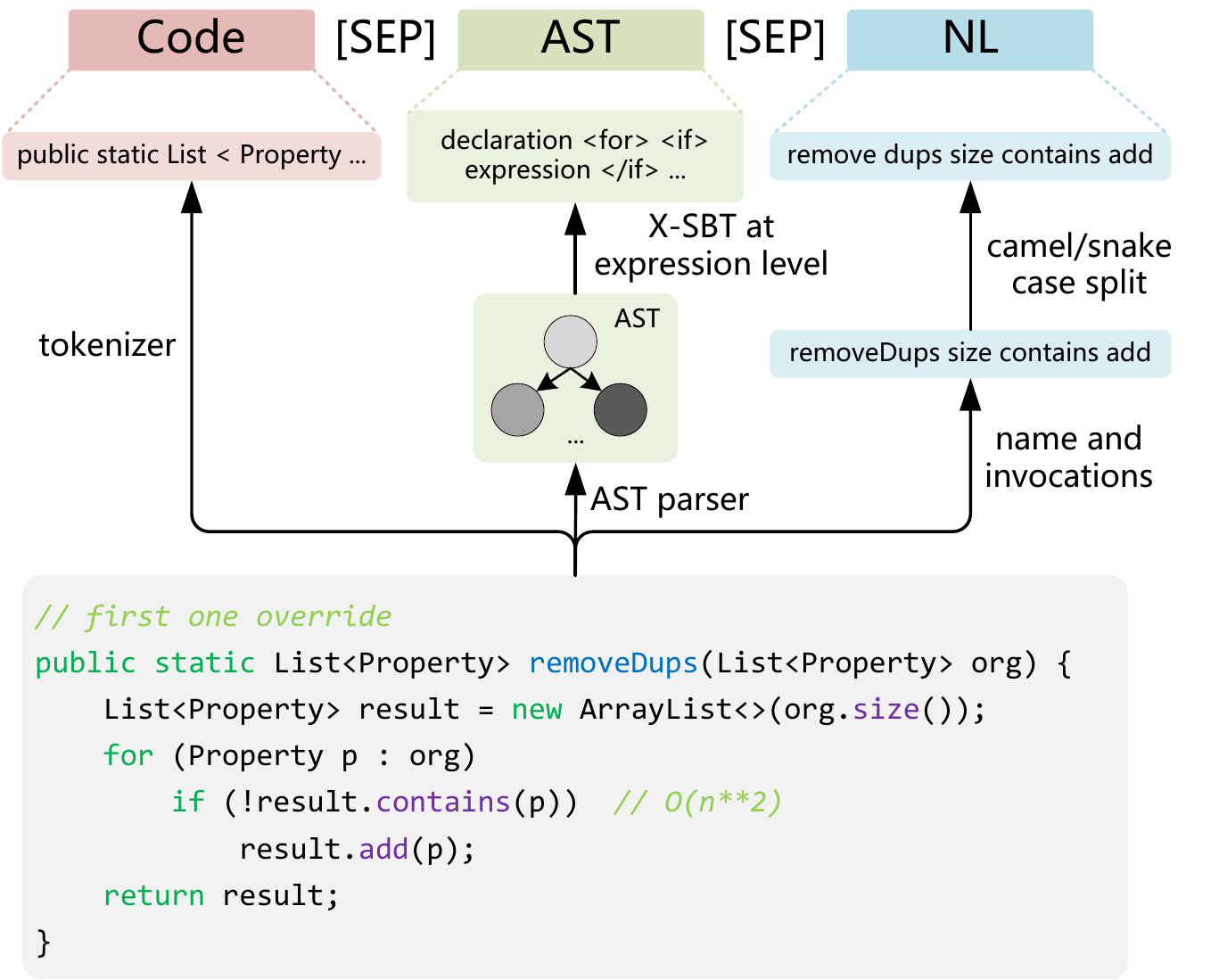}
  \caption{The input of a real world Java code snippet. Due to space constraints, we abbreviated the node names in the AST sequence. ``NL'' denotes the natural language input.}
  \Description{Input of a Java method.}
  \label{figure:model_input}
\end{figure}

The inputs of the model are three different types of components belonging to a complete \textit{Method} (i.e., it can be invoked by its name), namely, the code tokens, the linearized AST, and the natural language. In this subsection, we will demonstrate this with a real Java method\footnote{https://github.com/Unidata/thredds/blob/d2d68f9eee87f345625211324d71d5dc3e162\\ee1/cdm/src/main/java/thredds/client/catalog/Property.java\#L56-L63} shown in the bottom of Figure~\ref{figure:model_input}.

\subsubsection{Code Tokens}
\label{section:code_tokens}

As we can see from Figure~\ref{figure:model_input}, the first part of the input is the code token sequence of a method. We use a lexical analyzer to tokenize the source code and then obtain the tokens $C=\{c_1,c_2,\dots,c_l\}$, where $l$ is the number of code tokens. Specifically, we use the Python standard library\footnote{https://docs.python.org/3.8/library/tokenize.html} to tokenize Python codes. For languages such as Java, JavaScript, PHP and Go, we use the Python binding\footnote{https://pypi.org/project/antlr4-python3-runtime/} of ANTLR 4\footnote{https://github.com/antlr/antlr4} to get the code tokens. The Ruby source code is tokenized by the calling of a Ruby binary program. The source codes of other programming languages are tokenized by the NLTK tokenizer\footnote{https://www.nltk.org/api/nltk.tokenize.html}.

\subsubsection{Linearized AST}
\label{section:linearized_ast}

To represent the second part of the input, i.e., the structural information of the code, we convert an AST into a specially formatted sequence by traversing it, and call the result of this converting a \textit{linearized} AST. We first employ an AST parser\footnote{https://tree-sitter.github.io/tree-sitter/} to get the corresponding AST, then use a traversal method to parse the AST into a sequence. 

\begin{figure}[htbp]
  \centering
  \includegraphics[width=0.45\textwidth]{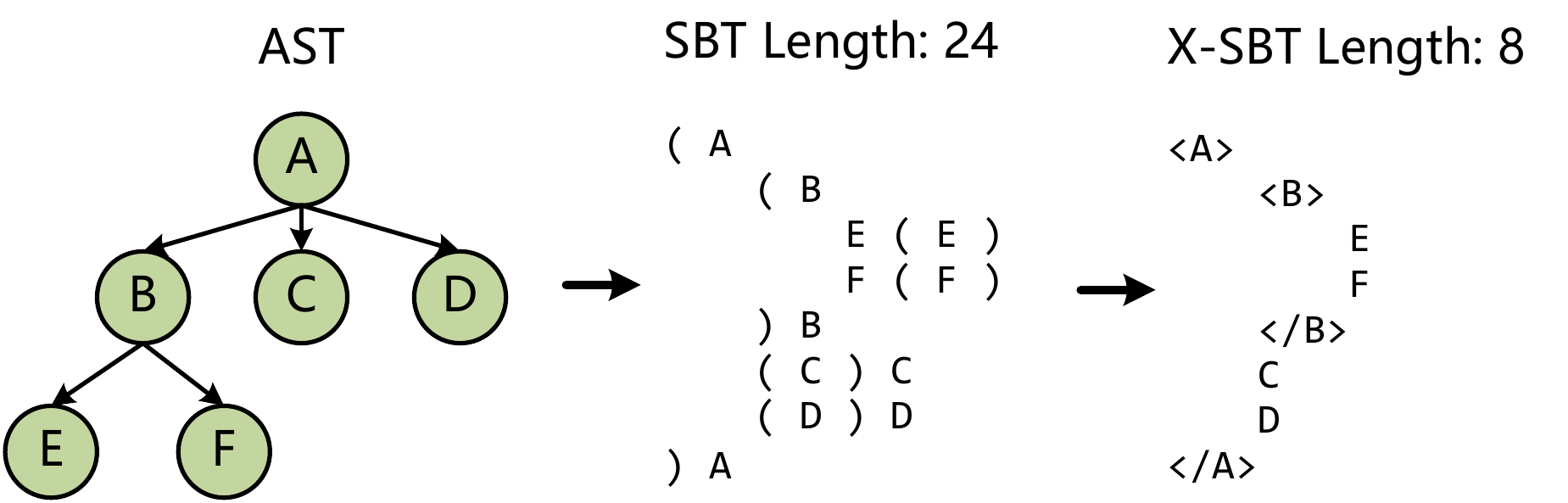}
  \caption{An example of the SBT and X-SBT, traversal sequences are formatted and indented for a better illustration.}
  \Description{Comparison of SBT and X-SBT.}
  \label{figure:xsbt}
\end{figure}

Instead of using the original SBT (please refer to~\cite{hu2018deep,hu2020deep} for more details), which has been shown to be more effective than classical traversal methods (i.e., pre-order traversal) but tend to produce excessively long sequences that are on average more than three times the length of the code, we propose a simplified version of SBT called X-SBT (XML-like SBT) to traverse ASTs. X-SBT can reduce the length of the resulting sequence of traversals by more than half, Figure~\ref{figure:xsbt}
shows a comparison of SBT and X-SBT. It can be seen that when traversing AST, SBT takes two tokens, ``('' and the node name, as the starting flag of a certain AST node, and takes ``)'' and the name as the ending flag. We make one observation: for non-terminal nodes (i.e., nodes which are not leaves), we can replace the starting flag with one token in an XML-like form, similarly for the ending flag. For terminal nodes, i.e., leaf nodes, we can further merge the starting and ending tokens into one token. Therefore, it is easy to prove that X-SBT shortens the length by more than half.

\begin{figure}[htbp]
  \centering
  \includegraphics[width=0.47\textwidth]{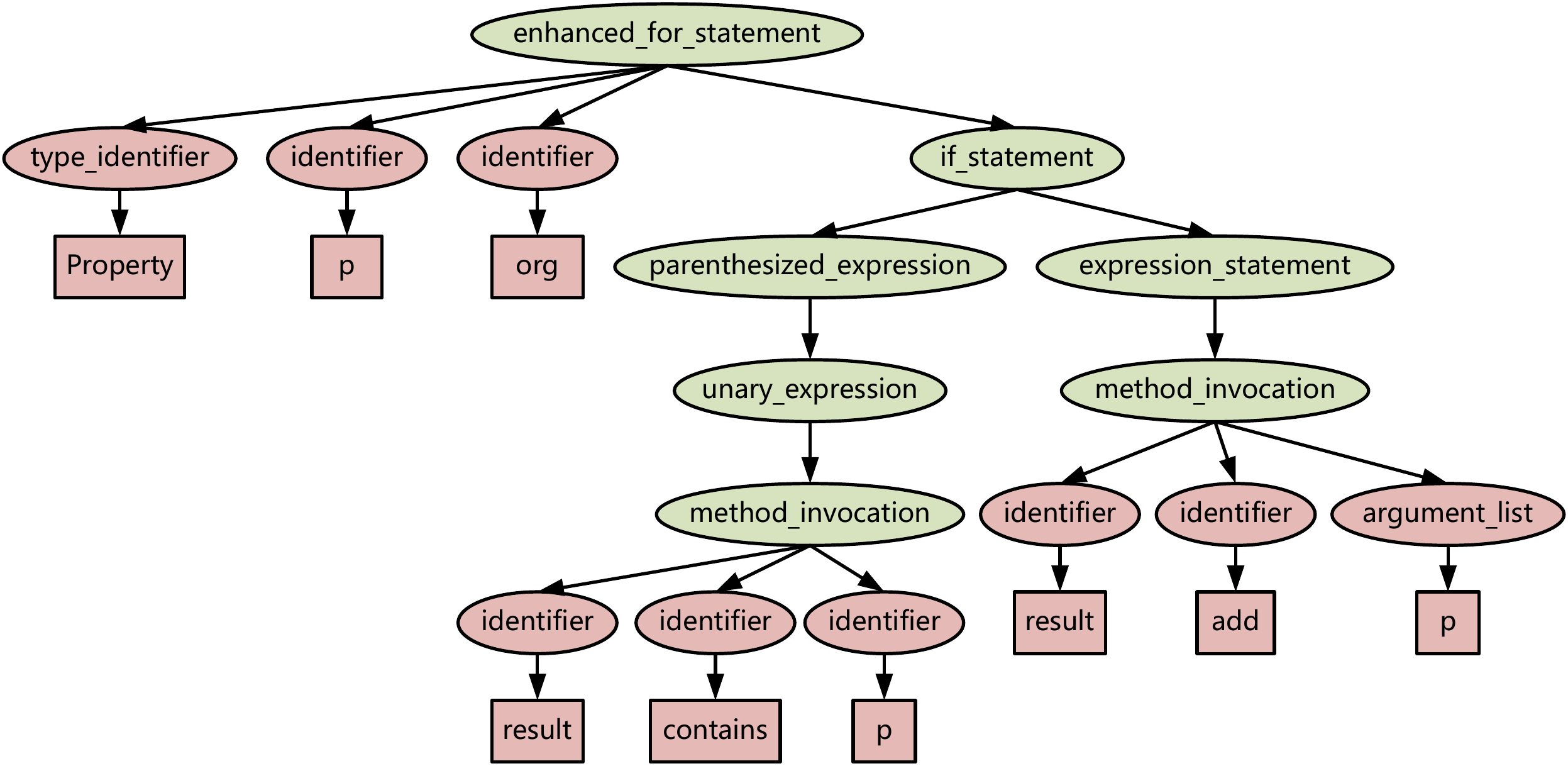}
  \caption{The AST of the ``for'' statement block of the Java code snippet in Figure~\ref{figure:model_input}. Non-terminal nodes are ellipses and terminal nodes are represented as rectangles. The green nodes will be traversed by X-SBT at the expression level, while the red ones will not.}
  \Description{Example of the AST.}
  \label{figure:xsbt_example}
\end{figure}

However, X-SBT sequences are still long. To further shorten X-SBT sequences, we make X-SBT traverse only the nodes at or above the \textit{expression} level in the AST. This will also reduce redundancy. Commonly, the AST contains both lexical and syntactic information, where the lexical information is already represented by the first part of the input (i.e., code tokens). Since the lexical information in the AST is concentrated on the terminal nodes, we only keep the nodes at or above the \textit{expression} level, so that it contains only syntactic information. Figure~\ref{figure:xsbt_example}
shows the nodes in the AST that will be traversed by the X-SBT at the \textit{expression} level. Obviously, it traverses only one subtree of the AST, and consequently, it can further reduce the length of the sequence by ignoring some fine-grained (lexical) information that is already present in the code tokens. To conclude, we traverse the AST using X-SBT at the \textit{expression} level to obtain the tokens $A=\{a_1,a_2,\dots,a_m\}$, where $m$ is the length of the sequence.

\subsubsection{Natural Language}
\label{section:nl_input}

For extracting natural language information from the code only, we derive the method name and API call sequence of the code\footnote{The reason we do not use documentation in code, such as docstring and in-line comments, is that documentation is not always available as we mentioned in Section~\ref{section:introduction}.}. For example, the tokens extracted from the code snippet in Figure~\ref{figure:model_input} are \texttt{removeDups, size, contains, add}. We further split each token of the form \textit{CamelCase} and \textit{snake\_case}, so \texttt{removeDups} is split into \texttt{remove} and \texttt{dups}. Then we take the resulting linear sequence of tokens, $N=\{n_1,n_2,\dots,n_p\}$, as our natural language input, where $p$ is the number of tokens.

After completing the construction of the three input parts, we concatenate them and use a special symbol, i.e., ``[SEP]'', to separate the three inputs. Therefore, the input is represented as
\begin{equation}
    Input=c_1,\dots,c_l,[\textup{SEP}],a_2,\dots,a_m,[\textup{SEP}],n_1,\dots,n_p
\end{equation}

\subsection{Pre-Training Tasks}
\label{section:pre_training_tasks}

In this section, we introduce the three tasks in the order they are used for (sequential) pre-training.

\subsubsection{Code-AST Prediction}
\label{section:cap}

The first pre-training task is Code-AST Prediction (CAP). Since we add structural information to the input, in CAP, we expect to have the model learn such information about the structure represented by X-SBT. So we introduce this binarized task that can be trivially generated from any code. Specifically, when we construct the input representation, namely $Input=C$,[SEP],$A$,[SEP],$N$, 50\% of the time $A$ is the actual AST sequence corresponding to $C$ (labeled as \texttt{IsAST}), and 50\% of the time it is a random AST sequence from the dataset (labeled as \texttt{NotAST}). As we show in Figure~\ref{figure:model_cls}, ``label'' is used for Code-AST Prediction. Note that CAP is closely similar to the next sentence prediction task~\cite{devlin2019bert}.

\subsubsection{MASS}
\label{section:mass}

Since SPT-Code is based on the encoder-decoder architecture, we expect to pre-train SPT-Code in a sequence-to-sequence style. Therefore, we then adopt MASS, which seeks to reconstruct a sentence fragment given the remaining part of the sentence in the encoder-decoder framework. We employ a modified version of MASS as one of our pre-training tasks, with the intention of training the model to understand, infer and generate code sequences.

Given an input $Input=C$,[SEP],$A$,[SEP],$N$, we denote $C^{\backslash u:v}_{\textup{origin}}$ as a modified version of $C$ where its fragment from position $u$ to $v$ is masked, that is,
\begin{equation}
  C^{\backslash u:v}_{\textup{origin}}=\{c_0,\dots,c_{u-1},[MASK],\dots,[MASK],c_{v+1},\dots,c_l\}
\end{equation}
where $0<u<v<l$. In contrast, $C^{u:v}$ denotes the fragment of $C$ from $u$ to $v$. In our work, we merge a number of consecutive ``[MASK]'' symbols into one, then $C^{\backslash u:v}_{\textup{origin}}$ is replaced by $C^{\backslash u:v}$:
\begin{equation}
  C^{\backslash u:v}=\{c_0,\dots,c_{u-1},[MASK],c_{v+1},\dots,c_l\}
\end{equation}

We utilize this modified version of MASS to pre-train our sequence to sequence model by predicting the fragment $C^{u:v}$ taking the input $Input=C^{\backslash u:v}$,[SEP],$A$,[SEP],$N$. Then, $k=u-v+1$ is the number of the masked consecutive tokens, and we follow Song et al.~\cite{song2019mass} and set $k$=50\% of $l$ to achieve the best performance.

\subsubsection{Method Name Generation}
\label{section:mng}

With the last pre-training task, we would expect the model to learn information such as the intent and functionality of the code. The method name, which is present in every method, can be seen as an extremely concise summary of the method. Xie et al.~\cite{xie2021exploiting} analyze the method name and code summary in a Java dataset built by Leclair and McMillan~\cite{leclair2019recommendations}, and find that on average 50.6\% of the words in method names appear in the corresponding summaries, and 21.3\% of the words in the summaries appear in the corresponding method names. For about 20\% of the methods, all the words in the method names appear in the corresponding summaries. Therefore, they improve code summarization performance successfully by predicting method names in advance.

Therefore, we exploit method names in this pre-training task, which we call Method Name Generation (MNG). Given the model input $Input=C$,[SEP],$A$,[SEP],$N$, we denote $name=c_i$ as the method name in $C$, where $i$ is the index of the method name token in $C$. From Section~\ref{section:nl_input}, we learn that the split $name=c_i$ also appears in $N$, so we remove the tokens from $N$ that are derived from the method name split. Assuming that the method name is split into $s$ subtokens, the final input of MNG will be as follows:
\begin{equation}
  \begin{split}
    Input_{MNG}=&c_0,\dots,c_{i-1},[MASK],c_{i+1},\dots,c_l,\\
    &[SEP],a_1,a_2,\dots,a_m,\\
    &[SEP],n_{s+1},n_{s+2},\dots,n_p
  \end{split}
\end{equation}
Then we make the decoder try to generate the split method name, i.e., $n_1,n_2,\dots,n_s$.

\subsection{Fine-Tuning}
\label{section:fine-tuning}

We group all downstream tasks into two categories: classification and generation. For classification tasks, we use the setting shown in Figure~\ref{figure:model_cls}; otherwise, we use the setting in Figure~\ref{figure:model_gen}. For each task, we simply plug in the task-specific inputs and outputs into SPT-Code and fine-tune all the parameters end-to-end. For example, in code search, which we cast as a classification task, when obtaining a representation vector of a natural language query, we take only the natural language as input and do not consider the first two parts of the input, i.e., code and AST. For code summarization, which we cast as a generation task, we only include the natural language summary in the output. We will describe the task-specific details in the corresponding subsections of Section~\ref{section:experiment}.

\section{Experiment}
\label{section:experiment}

In this section, we first introduce the pre-training data and settings (Section~\ref{section:exp_pre_train}) and how we fine-tune on the downstream tasks (Section~\ref{section:exp_fine_tune}). Then, we answer four research questions (Section~\ref{section:exp_evaluation}). Finally, we manually evaluate SPT-Code through quantitative and qualitative analysis.

\subsection{Pre-Training}
\label{section:exp_pre_train}

\subsubsection{Dataset}
\label{section:pre_training_dataset}

The dataset we use for pre-training SPT-Code is CodeSearchNet~\cite{husain2020codesearchnet}, which has also been used to pre-train CodeBERT~\cite{feng2020codebert}, GraphCodeBERT~\cite{guo2020graphcodebert} and T5-learning~\cite{mastropaolo2021studying}. The CodeSearchNet corpus is programmatically obtained by scraping open-source repositories and pairing individual functions with their (processed) documentation as natural language annotation. It includes more than 6.4M codes from 6 programming language, i.e., Java, Python, JavaScript, PHP, Go and Ruby. The data statistics are shown in Table~\ref{table:csn_statistics}.
\begin{table}[htbp]
  \centering
  \caption{Pre-Training Dataset Size Statistics}
  \label{table:csn_statistics}
  \small
  \begin{tabular}{lrr}
    \toprule
    \multirow{2}{*}{\textbf{Language}} & \multicolumn{2}{c}{\textbf{\# Methods/Function}} \\ \cline{2-3}
    & \multicolumn{1}{r}{\textbf{w/ documentation}} & \multicolumn{1}{r}{\textbf{All}} \\
    \midrule
    Java       & 542,991 & 1,569,889 \\
    Python     & 503,502 & 1,156,085 \\
    JavaScript & 157,988 & 1,857,835 \\
    PHP        & 717,313 & 977,821 \\
    GO         & 347,789 & 726,768 \\
    Ruby       & 57,393  & 164,048 \\
    \hline
    All        & 2,326,976 & 6,452,446 \\
    \bottomrule
  \end{tabular}
\end{table}

Since our input can be extracted from completely unlabeled data, we can make use of all the 6.4M data instances in CodeSearchNet. However, CodeBERT and GraphCodeBERT use labels (i.e., the documentation) in the input for training both code and natural language, so they are both pre-trained on all the data in the ``w/ documentation'' column of the ``All'' row, which we named \textit{CSN$_{w/ doc}$}. T5-learning, inputs the code and the documentation into the model separately when pre-training, so it resembles SPT-Code in the sense that it does not utilize the code-document correspondence. Nevertheless, the publicly available implementation of T5-learning uses only the data in the ``All'' column of ``Java'' row, named \textit{CSN$_{Java}$}.

\subsubsection{Settings}
\label{section:pre_train_settings}

The three pre-training tasks of SPT-Code are performed sequentially. We tried different orders and different numbers of epoch for these tasks and found that better results can be achieved by first pre-training CAP for 10 epochs, then MASS for 30 epochs, and eventually MNG for 30 epochs. All three tasks use cross-entropy as the loss function. We use AdamW~\cite{loshchilov2018decoupled} as our optimizer, the initial learning rate is 5e-5 and the warmup step is set as 2000. We pre-train SPT-Code on 4$\times$NIVDIA A100s\footnote{The GPUs are provided by Alibaba Group, which employ EFLOPS~\cite{eflops} and Alibaba Collective Communication Library (ACCL)~\cite{accl} techniques. EFLOPS is a high performance distributed training system that can achieve near-linear scalability of overall throughput, and ACCL brings the performant efficiency of EFLOPS architecture to general cluster systems and Cloud scenarios, which is able to achieve fully non-congested communication.} with a total batch size of 256.

We use Byte-Pair Encoding (BPE)~\cite{sennrich2016neural} to tokenize the code and the natural language (after CamelCase and snake\_case splitting), and use regular word tokenizer to tokenize X-SBT sequences. The tokenizer is built upon the whole pre-training data, and will be directly used on each downstream tasks without any modification.

\subsection{Fine-Tuning on Downstream Tasks}
\label{section:exp_fine_tune}

In this subsection, we detail the process of fine-tuning the pre-trained SPT-Code on the five downstream tasks. For each downstream task, we give a brief introduction, tell the datasets for fine-tuning, present the compared baselines, and eventually illustrate the metrics for evaluating.

\subsubsection{Code Summarization}
\label{section:fine_tune_summarization}

Code summarization, also known as the process of code summary or code comment generation, is the task of automatically generating a natural language description for a piece of source code that summarizes the overall actions of the code snippet accurately, adequately, and concisely~\cite{sridhara2010towards}. Work in this area of research has generally focused on generating a (typically short) natural language comment from a given method.

\textbf{Datasets.} In addition to CodeSearchNet, we use two widely used classical datasets that are collected from open source repositories in GitHub, i.e., the Java dataset (JCSD)~\cite{hu2018summarizing} and the Python dataset (PCSD)~\cite{miceli2017parallel}. As for JCSD, the comment of each method is the first sentence extracted from its Javadoc, and the comment of a method in PCSD is its docstring.

\textbf{Baselines.} We adopt CodeBERT~\cite{feng2020codebert}, GraphCodeBERT~\cite{guo2020graphcodebert}, CugLM~\cite{liu2020multi}, T5-learning~\cite{mastropaolo2021studying} and TreeBERT~\cite{jiang2021treebert} as baselines for all datasets
\footnote{We resize all the pre-trained models here and in all the following downstream tasks identical to ours.}. For JCSD, PCSD and CodeSearchNet's Java/Python datasets, we adopt NerualCodeSum~\cite{ahmad2020transformer}, which employs Relative Position~\cite{shaw2018self} and Copy Attention~\cite{see2017get} upon vanilla Transformer~\cite{vaswani2017attention}, as another baseline\footnote{We have tried to run it as Java for other languages in CodeSearchNet but got completely nonsensical output.}. Besides, we use three more baselines for JCSD and PCSD, namely, DeepCom~\cite{hu2018deep}, Rencos~\cite{zhang2020retrieval}, and SiT~\cite{wu2021code}, which are widely compared or recently proposed RNN or transformer-based models dedicated for code summarization.

\textbf{Metrics.} We use BLEU (B.)~\cite{papineni2002bleu}, METEOR (M.)~\cite{banerjee2005meteor} and ROUGE-L (R.L)~\cite{lin2002manual}, which are widely used in fields such as machine translation and text summarization, to measure the similarity between the sentences generated by the model and the goal sentences.

\subsubsection{Code completion}
\label{section:fine_tune_completion}

Code completion is to generate code given its surrounding code as context. We fine-tune SPT-Code on a task called \textit{any-code} completion~\cite{alon2020structural} here. Unlike \textit{restricted} completion where the target code contains only primitive types (e.g., \texttt{int} and \texttt{string}) and excludes user-defined functions, \textit{any-code} completion aims to generate code in a general-purpose programming language without any restriction on its vocabulary or structure. Specifically, consider a program $\mathcal{P}$ and some part of the program $p\in \mathcal{P}$, any-code completion makes the model to predict $p$ given the rest of the program $\mathcal{P}^-=\mathcal{P}/p$.

\textbf{Datasets.} The dataset we use for any-code completion is the Java dataset provided by Alon et al.~\cite{alon2020structural}. It is based on the Java-small dataset~\cite{alon2018code2seq}, which contains the least code duplication~\cite{allamanis2019adverse}. Alon et al. create any-code completion examples by selecting every expression larger than a single AST node as the target $p$, using the reminder of the method as the context $\mathcal{P}$. They also filter many methods to clean the dataset and make the task harder. The resulting dataset contains 1.3M/10k/20k train/dev/test examples.

\textbf{Baselines.} Besides CodeBERT, GraphCodeBERT, CugLM, T5-learning and TreeBERT, some specific methods for this task are also compared. Code2seq~\cite{alon2018code2seq} represents a code snippet as the set of compositional paths in its AST and uses attention to select the relevant paths while decoding. Transformer w/ copy uses the implementation of Open-NMT~\cite{klein2017opennmt} with a copy mechanism~\cite{gu2016incorporating}. Seq2tree w/ copy~\cite{aharoni2017towards} learns to generate the linearized, subtokenized target AST for code completion. SLM~\cite{alon2020structural} leverages joint modeling of an AST and its missing subtree using a structural language model, which estimates the probability of the program’s AST by decomposing it into a product of conditional probabilities over its nodes. 

\textbf{Metrics.} Following Alon et al.~\cite{alon2020structural}, we use exact match accuracy at 1 (Acc@1) and 5 (Acc@5) for evaluation. An exact match is counted if and only if the sentence generated by the model is identical to the goal sentence (ignoring cases and whitespaces). If we let the model generate $k$ candidate sentences with the highest probability and if an exact match is counted while any one of the candidates matches the goal sentence exactly, the ultimate exact match accuracy is Acc@$k$ (e.g., if only one candidate, it is Acc@1). 

\subsubsection{Bug fixing}
\label{section:fine_tune_bug_fix}

Bug fixing, aka. code repair or code refinement, aims to automatically fix bugs in the code. It can help software developers locate and fix bugs and thus save a lot of time~\cite{jorgensen2006systematic,tufano2019empirical}.

\textbf{Datasets.} We use the dataset collected by Tufano et al.~\cite{tufano2019empirical}, who extract method-level pairs of buggy and corresponding fixed code named BFPs (bug-fix pairs) from bug-fixing commits in thousands of GitHub Java repositories. Each BFP is composed of a tuple $BFP=<m_b, m_f>$, where $m_b$ represents a buggy code component, $m_f$ represents the corresponding fixed code. Based on the code size, Tufano et al. provide two datasets: BFP$_{\textup{small}}$ and BFP$_{\textup{medium}}$, with the former having a code length below 50 and the latter having a code length between 50 and 100. 

\textbf{Baselines.} The original method proposed by the dataset creator~\cite{tufano2019empirical} is marked as \textit{Tufano et al.} S2S+COPYSAPN~\cite{panthaplackel2021copy} is an extension of seq2seq models which can copy entire spans of the input to the output in one step and reduce the number of decisions required during inference. CodeBERT, GraphCodeBERT, CugLM, T5-learning and TreeBERT are also used as baselines. 

\textbf{Metrics.} We report Acc (=Acc@1) and BLEU on both datasets. Both $m_b$ and $m_f$ are abstracted (please refer to \cite{tufano2019empirical} for more details) before being fed to the model, and we do not translate the abstracted code predicted by the model back into source code before evaluation using the metrics because the result before and after translation are the same.

\subsubsection{Code Translation}
\label{section:fine_tune_translation}

Code translation is important for migrating legacy code in one language into an ecosystem built in a different language.

\textbf{Datasets.} Following Chen et al.~\cite{chen2018tree} and Guo et al.~\cite{guo2020graphcodebert}, we conduct our experiments on a dataset containing several open-source projects, which have both a Java and a C\# implementation.

\textbf{Baselines.} Baselines are Naive (i.e., directly copying the source code as the translation result), vanilla Transformer~\cite{vaswani2017attention}, CodeBERT, GraphCodeBERT, CugLM, T5-learning and TreeBERT. 

\textbf{Metrics.} We use Acc and BLEU as metrics.

\subsubsection{Code Search}
\label{section:fine_tune_search}

Code search aims to find the code snippet that most closely resembles the semantics of the given natural language query statement from a set of codes, which is named \textit{codebase} in this task. It is a good choice for testing the performance of SPT-Code in classification mode. For each code-query pair $(c,q)$, we compute the representation vectors of $c$ and $q$, i.e., $V_c$ and $V_q$ respectively. Then, we randomly take another query statement from the dataset as a negative sample, denoted as $q-$, whose representation vector is $V_{q-}$. To ensure the effectiveness of the negative samples, we restrict the BLEU score~\cite{papineni2002bleu} between $q-$ and $q$ to be lower than 0.3. The training loss is:
\begin{gather}
  \mathcal{L}_{search}=\sum_{(c,q,q-)\in D} \max (0, \varepsilon -\cos (V_c,V_q)+\cos (V_c,V_{q-}))\\
  \cos (V_c,V_q)=\frac{V_c\cdot V_q}{\Vert V_c\Vert \Vert V_q\Vert }
\end{gather}
where $D$ is the dataset, $\varepsilon$ is a fixed value of 0.05 (following Gu et al.~\cite{gu2018deep}), and $\cos$ denotes the calculation of cosine similarity. In the evaluation, we first get the representation vectors of all codes in codebase as candidates. Then, for each vector of query statements, we select a few code representation vectors from the candidates that are closest to the vector of query statements. Finally, the metric is calculated.

\textbf{Datasets.} We use CodeSearchNet~\cite{husain2020codesearchnet}. 

\textbf{Baselines.} We use CNN~\cite{kim2014convolutional} (i.e., 1D convolutional neural network), Bi-GRU~\cite{cho2014learning}, and Transformer (i.e., vanilla Transformer~\cite{vaswani2017attention}) as baselines in addition to CodeBERT and GraphCodeBERT. 

\textbf{Metrics.} We use Mean Reciprocal Rank (MRR) for evaluation. MRR is a statistic measure for evaluating search algorithms. The reciprocal rank of a query response is the multiplicative inverse of the rank of the first correct answer: 1 for first place, 1/2 for second place. Therefore, the mean reciprocal rank is the average of the reciprocal ranks of results for a sample of queries Q:
\begin{equation}
    \textup{MRR}=\frac{1}{|Q|}\sum_{i=1}^{|Q|}\frac{1}{\textup{rank}_i}
\end{equation}

\subsection{Evaluation}
\label{section:exp_evaluation}

We evaluate SPT-Code by answering four research questions.

\subsubsection*{RQ1: How effective is SPT-Code compared with the state-of-the-art baselines and other code pre-trained models on five downstream tasks?}
\label{section:rq_comparsion}

We conduct experiments on five downstream tasks introduced in the previous subsection. The results on classical and CodeSearchNet code summarization are shown in Tables~\ref{table:results_summarization}
\begin{table}[htbp]
  \centering
  \caption{Results on classical code summarization.}
  \label{table:results_summarization}
  \small
  \begin{tabular}{lcccccc}
    \toprule
        \multicolumn{1}{l}{\multirow{2}{*}{Methods}} &
        \multicolumn{3}{c}{JCSD} &
        \multicolumn{3}{c}{PCSD} \\ \cline{2-7}
        \multicolumn{1}{c}{} &
        \multicolumn{1}{c}{B.} &
        \multicolumn{1}{c}{M.} &
        \multicolumn{1}{c}{R.L} &
        \multicolumn{1}{c}{B.} &
        \multicolumn{1}{c}{M.} &
        \multicolumn{1}{c}{R.L} \\
    \midrule
        DeepCom       & 37.5 & 22.0 & 51.2 & 20.4 &  9.5 & 36.8 \\
        Rencos        & 36.9 & 26.5 & 51.2 & 28.9 & 20.4 & 45.0 \\
        NerualCodeSum & 44.5 & 26.4 & 54.7 & 32.3 & 19.7 & 46.8 \\
        SiT           & 45.3 & 27.3 & 55.0 & 33.8 & 20.7 & 48.2 \\
    \hline
        CodeBERT      & 45.2 & 26.2 & 54.7 & 33.3 & 21.6 & 49.2 \\
        GraphCodeBERT & 46.0 & 26.5 & 56.5 & 33.9 & 22.1 & 50.4 \\
        CugLM         & 46.1 & 26.3 & 55.8 & 34.3 & 21.6 & 49.7 \\
        T5-learning   & 44.2 & 26.5 & 53.9 & 34.1 & 22.6 & 50.1 \\
        TreeBERT      & 47.9 & 27.2 & 56.6 & 34.7 & 23.0 & 50.5 \\
    \hline
        SPT-Code & \textbf{49.1} & \textbf{32.4} & \textbf{58.2} & \textbf{36.1} & \textbf{26.9} & \textbf{52.0} \\
    \bottomrule
  \end{tabular}
\end{table}
and Table~\ref{table:results_summarization_csn},
\begin{table*}[htbp]
  \centering
  \caption{Results on CodeSearchNet code summarization.}
  \label{table:results_summarization_csn}
  \small
    \begin{tabular}{lcccccccccccccccccc}
    \toprule
      \multirow{2}{*}{Methods} & 
      \multicolumn{3}{c}{Java} & 
      \multicolumn{3}{c}{Python} & 
      \multicolumn{3}{c}{JavaScript} &
      \multicolumn{3}{c}{PHP} & 
      \multicolumn{3}{c}{Go} & 
      \multicolumn{3}{c}{Ruby} \\
    \cline{2-19} &
      B. & M. & R.L & 
      B. & M. & R.L & 
      B. & M. & R.L &
      B. & M. & R.L & 
      B. & M. & R.L & 
      B. & M. & R.L \\
    \midrule
      NeuralCodeSum & 10.7 & 12.9 & 25.9 & 9.4  &  4.9 & 17.0 & - & - & - & - & - & - & - & - & - & - & - & - \\
      CodeBERT      & 13.6 & 16.4 & 32.2 & 10.7 &  6.0 & 21.0 & 10.0 & 10.5 & 22.2 & 20.1 & 19.3 & 42.7 & 16.0 & 11.7 & 30.4 &  8.3 &  7.5 & 17.2 \\
      GraphCodeBERT & 14.5 & 17.8 & 33.5 & 11.0 &  7.4 & 22.1 & 10.9 & 11.4 & 24.9 & 20.0 & 20.2 & 42.8 & 16.5 & 12.5 & 30.1 &  8.3 &  8.1 & 18.0 \\
      CugLM         & 13.2 & 16.1 & 31.7 & 11.4 &  7.5 & 22.0 & 10.0 & 10.4 & 21.8 & 17.8 & 18.2 & 41.9 & 17.0 & 12.7 & 29.7 &  7.4 &  6.8 & 15.3 \\
      T5-learning   & 13.5 & 16.8 & 34.2 &  9.7 &  5.5 & 16.3 &  9.0 &  9.2 & 19.6 & 18.4 & 17.9 & 35.6 & 15.3 & 11.0 & 25.9 &  7.8 &  6.7 & 15.5 \\
      TreeBERT      & 13.8 & 17.1 & 34.3 & 11.1 &  7.2 & 21.6 & 10.3 & 10.5 & 22.0 & 18.0 & 19.1 & 42.4 & 17.1 & 12.8 & 30.1 &  7.4 &  6.9 & 15.5 \\
    \hline
      SPT-Code & \textbf{16.8} & \textbf{20.6} & \textbf{36.3} & \textbf{12.8} & \textbf{14.2} & \textbf{27.1} & \textbf{12.8} & \textbf{17.2} & \textbf{28.8} & \textbf{20.4} & \textbf{20.6} & \textbf{43.7} & \textbf{18.8} & \textbf{16.6} & \textbf{38.1} & \textbf{8.4} & \textbf{11.1} & \textbf{20.0} \\
    \bottomrule
    \end{tabular}
\end{table*}
respectively. Tables~\ref{table:results_completion}-\ref{table:results_search} show the results of code completion, bug fixing, code translation and code search, respectively.
\begin{table}[htbp]
  \centering
  \caption{Results on any-code completion.}
  \label{table:results_completion}
  \small
  \begin{tabular}{lcc}
    \toprule
      Methods & Acc@1 & Acc@5 \\
    \midrule
      code2seq            & 10.44 & 15.47 \\
      Transformer w/ copy & 16.68 & 24.12 \\
      seq2tree w/ copy    & 16.46 & 22.89 \\
      SLM                 & 18.00 & 24.77 \\
    \hline
      CodeBERT            & 17.23 & 24.79 \\
      GraphCodeBERT       & 18.21 & 25.00 \\
      CugLM               & 18.43 & 25.51 \\
      T5-learning         & 16.03 & 23.78 \\
      TreeBERT            & 17.17 & 24.73 \\
    \hline
      SPT-Code            & \textbf{19.09} & \textbf{26.57} \\
    \bottomrule
  \end{tabular}
\end{table}
\begin{table}
  \centering
  \caption{Results on bug fixing.}
  \label{table:results_bug_fix}
  \small
  \begin{tabular}{lcccc}
    \toprule
      \multicolumn{1}{l}{\multirow{2}{*}{Methods}} &
      \multicolumn{2}{c}{BFP$_{\textup{small}}$} &
      \multicolumn{2}{c}{BFP$_{\textup{medium}}$} \\ \cline{2-5}
      \multicolumn{1}{c}{} &
      \multicolumn{1}{c}{Acc} &
      \multicolumn{1}{c}{BLEU} &
      \multicolumn{1}{c}{Acc} &
      \multicolumn{1}{c}{BLEU} \\
    \midrule
      Tufano et al. & 9.27  & -     &  3.21 & - \\
      S2S+COPYSPAN  & \textbf{17.6} & -     & 7.9  & - \\
    \hline
      CodeBERT      &  6.23 & 61.47 &  2.43 & 68.54 \\
      GraphCodeBERT &  7.49 & 63.40 &  3.16 & 67.33 \\
      CugLM         & 14.55 & 67.24 &  7.84 & 71.23 \\
      T5-learning   & 12.01 & 65.02 &  4.51 & 60.89 \\
      TreeBERT      & 13.15 & 65.74 &  7.61 & 70.67 \\
    \hline
      SPT-Code & 17.54 & \textbf{75.10} & \textbf{10.86} & \textbf{87.88} \\
    \bottomrule
  \end{tabular}
\end{table}
\begin{table}
  \centering
  \caption{Results on code translation.}
  \label{table:results_translation}
  \small
  \begin{tabular}{lcccc}
    \toprule
      \multicolumn{1}{l}{\multirow{2}{*}{Methods}} &
      \multicolumn{2}{c}{Java $\to$ C\#} &
      \multicolumn{2}{c}{C\# $\to$ Java} \\ \cline{2-5}
      \multicolumn{1}{c}{} &
      \multicolumn{1}{c}{Acc} &
      \multicolumn{1}{c}{BLEU} &
      \multicolumn{1}{c}{Acc} &
      \multicolumn{1}{c}{BLEU} \\
    \midrule
      Naive         & 00.0 & 18.52 & 00.0 & 18.66 \\
      Transformer   & 33.4 & 56.05 & 38.8 & 50.84 \\
    \hline
      CodeBERT      & 58.8 & 79.34 & 57.3 & 71.10 \\
      GraphCodeBERT & 59.3 & 80.33 & 58.1 & 72.39 \\
      CugLM         & 61.6 & 82.72 & 59.5 & 74.89 \\
      T5-learning   & 54.1 & 81.35 & 48.4 & 77.10 \\
      TreeBERT      & 60.0 & 80.92 & 57.8 & 77.15 \\
    \hline
      SPT-Code & \textbf{64.07} & \textbf{90.34} & \textbf{60.29} & \textbf{86.10} \\
    \bottomrule
  \end{tabular}
\end{table}
\begin{table}[htbp]
  \centering
  \caption{Results on code search.}
  \label{table:results_search}
  \small
  \begin{tabular}{lcccccc}
    \toprule
      Methods & Java & Python & JS & PHP & Go & Ruby \\
    \midrule
      CNN           & 0.262 & 0.241 & 0.224 & 0.261 & 0.676 & 0.274 \\
      Bi-GRU        & 0.312 & 0.298 & 0.195 & 0.340 & 0.691 & 0.214 \\
      Transformer   & 0.404 & 0.399 & 0.289 & 0.430 & 0.729 & 0.278 \\
    \hline
      CodeBERT      & 0.673 & 0.670 & 0.618 & 0.624 & 0.879 & 0.674 \\
      GraphCodeBERT & 0.690 & 0.692 & \textbf{0.643} & 0.647 & \textbf{0.896} & \textbf{0.701} \\
    \hline
      SPT-Code      & \textbf{0.700} & \textbf{0.699} & 0.641 & \textbf{0.651} & 0.895 & \textbf{0.701} \\
    \bottomrule
  \end{tabular}
\end{table}

First of all, we can see that whether the baselines are dedicated to a specific downstream task or pre-trained models, SPT-Code clearly outperforms them in the vast majority of cases, including code summarization, code completion, bug fixing on BFP$_{\textup{medium}}$, and code translation. This is because during the pre-training process, SPT-Code learns the representation of codes from a large amount of data, as well as the connection between codes and structural and natural language information. In addition, the model is enhanced to generate code and natural language sequences by pre-training.

As for bug fixing on BFP$_{\textup{small}}$, the accuracy of SPT-Code is slightly lower than that of S2S+COPYSPAN, which achieves the best accuracy on BFP$_{\textup{small}}$. The reason we believe is two-fold. The first is the dataset, where the lengths of the code are up to 50, which allows RNN-based S2S+COPYSPAN to adequately cope with inputs of these lengths. Another reason is that the mechanism proposed by S2S+COPYSPAN to copy the entire span of the input is well suited for a task like bug fixing, where only individual modifications are made on the input. However, in the next RQ, we will show that SPT-Code still outperforms S2S+COPYSPAN on BFP$_{\textup{small}}$ after removing MNG from the pre-training tasks.

Finally, we can learn that SPT-Code performs comparably to GraphCodeBERT on code search, which suggests that although we designed our model to perform better on generation tasks, it does not come at the cost of the performance on classification tasks.

\subsubsection*{RQ2: How do the three pre-training tasks, as well as the AST and the natural language input contribute to the model's performance on the different downstream tasks?}
\label{section:rq_component}

In order to find the impact of each component to the overall performance of SPT-Code, we conduct ablation study on all downstream tasks. We remove each/all pre-training task, and AST or/and natural language part from the input, respectively. Owing to space limitations, for code summarization, we only show results on Java and Python in CodeSearchNet\footnote{Java and Python represent static and dynamic languages, respectively.}; and for code search, we only show results on Java and Go\footnote{The results for both are above and below GraphCodeBERT, respectively.}. The results are shown in the first three groups of Table~\ref{table:ablation_mixed}.
\begin{table*}
    \centering
    \caption{Ablation study on downstream tasks.}
    \label{table:ablation_mixed}
    \resizebox{\textwidth}{!}{
    \begin{tabular}{lcccccc|cc|cccc|cccc|cc}
        \toprule
            \multirow{3}{*}{Methods} &
            \multicolumn{6}{c|}{Summarization} &
            \multicolumn{2}{c|}{Completion} &
            \multicolumn{4}{c|}{Bug fixing} & 
            \multicolumn{4}{c|}{Translation} &
            \multicolumn{2}{c}{Search}\\ \cline{2-19}
            &
            \multicolumn{3}{c}{Java} &
            \multicolumn{3}{c|}{Python} &
            \multirow{2}{*}{Acc@1} &
            \multirow{2}{*}{Acc@5} &
            \multicolumn{2}{c}{BFP$_\textup{small}$} &
            \multicolumn{2}{c|}{BFP$_\textup{medium}$} &
            \multicolumn{2}{c}{Java $\to$ C\#} &
            \multicolumn{2}{c|}{C\# $\to$ Java} &
            \multirow{2}{*}{Java} &
            \multirow{2}{*}{Go} \\ \cline{2-7} \cline{10-17}
            & B. & M. & R.L & B. & M. & R.L & & & Acc & BLEU & Acc & BLEU & Acc & BLEU & Acc & BLEU & & \\
        \midrule
            SPT-Code & \textbf{16.79} & \textbf{20.55} & \textbf{36.34} & \textbf{12.77} & \textbf{14.16} & \textbf{27.10} & 19.09 & \textbf{26.57} & 17.54 & \textbf{75.10} & 10.86 & 87.77 & \textbf{64.07} & \textbf{90.34} & \textbf{60.29} & \textbf{86.10} & \textbf{0.700} & \textbf{0.895} \\
        \hline
            \ -w/o CAP  & 15.85 & 20.33 & 36.30 & 12.26 & 14.16 & 26.57 & 18.42 & 25.63 & 17.20 & 73.54 & 10.06 & 87.26 & 57.25 & 88.80 & 54.40 & 84.66 & 0.668 & 0.872 \\
            \ -w/o MASS & 15.75 & 20.19 & 35.70 & 12.16 & 13.48 & 26.00 & 17.42 & 23.71 & 16.30 & 72.71 & 9.65  & 86.10 & 53.93 & 82.09 & 50.55 & 82.32 & 0.686 & 0.889 \\
            \ -w/o MNG  & 15.45 & 19.86 & 35.65 & 12.01 & 13.81 & 25.98 & \textbf{19.71} & 26.10 & \textbf{18.11} & 73.09 & \textbf{11.96} & \textbf{87.79} & 58.80 & 88.79 & 57.16 & 85.19 & 0.673 & 0.875 \\
            \ -w/o all  & 13.54 & 17.72 & 33.31 & 11.90 & 12.74 & 25.23 & 13.49 & 19.31 & 14.28 & 72.89 & 8.27 & 85.30 & 49.19 & 79.28 & 47.49 & 78.69 & 0.656 & 0.855 \\
        \hline
            \ -w/o AST   & 16.43 & 20.14 & 36.19 & 11.87 & 13.80 & 25.74 & 18.53 & 25.78 & 16.35 & 72.85 & 10.61 & 87.52 & 55.81 & 87.79 & 56.17 & 82.66 & 0.692 & 0.885 \\
            \ -w/o NL    & 15.68 & 20.15 & 35.86 & 11.96 & 13.56 & 25.55 & 18.47 & 26.09 & 16.12 & 72.70 & 10.50 & 87.57 & 56.29 & 87.83 & 56.92 & 83.56 & 0.679 & 0.865 \\
            \ -only code & 15.35 & 19.97 & 35.63 & 10.25 & 12.74 & 23.90 & 18.33 & 25.88 & 16.01 & 72.55 & 10.19 & 86.96 & 54.13 & 87.34 & 54.59 & 83.81 & 0.669 & 0.864 \\
        \hline
            \ -CSN$_{\textup{w/ doc}}$ & 15.41 & 19.78 & 35.74 & 12.59 & 13.56 & 26.49 & 17.89 & 23.80 & 16.14 & 74.02 & 10.41 & 87.58 & 60.53 & 88.78 & 58.28 & 84.09 & 0.673 & 0.878 \\
            \ -CSN$_{\textup{Java}}$   & 15.67 & 19.72 & 35.22 & 12.40 & 12.60 & 25.24 & 17.76 & 23.54 & 16.10 & 74.51 & 10.18 & 86.96 & 59.43 & 86.55 & 58.11 & 83.07 & 0.678 & 0.866 \\
        \bottomrule
    \end{tabular}
    }
\end{table*}

The first thing we notice is that when we train from scratch (i.e., remove all pre-training tasks) or remove the AST and natural language from the input (i.e., only input code tokens), the performance of SPT-Code consistently drops considerably. Second, when we remove each of the three pre-training tasks, the results decreased in most cases, particularly in code summarization, code translation and code search. Of these, CAP is most useful for code search, MASS is most helpful for code translation, and removing MNG has the greatest impact on code summarization.

Interestingly, we find that for code completion and bug fixing, SPT-Code's performance w.r.t.\ accuracy improves instead when MNG is removed. This is understandable. On the one hand, code completion and bug fixing are tasks where the input and the output are both code, as is MASS, while the output of the MNG task is natural language, and thus the ability to generate natural language trained by MNG is not fully reflected in these two tasks. On the other hand, regarding code completion, MASS can be seen as totally unrestricted any-code completion. Both predict a piece of code based on its context, with the difference that in any-code completion, the piece of code is restricted to be an entire expression, while in MASS, the piece of code is selected completely at random. Therefore, fine-tuning code completion directly after MASS, i.e., removing the MNG, yields a higher result.

When we remove either the AST or the natural language from the input, the results of the model drop, indicating that they both help improve performance. In addition, for code summarization, bug fixing and code search, the results are lower when only natural language is removed, which indicates that in comparison, natural language helps these three tasks more than ASTs do. On the contrary, AST is more helpful for code completion and code translation.

Through ablation, we find that different pre-training tasks show different degrees of influence on the downstream tasks, As a result, appropriate trade-offs of pre-training tasks for different downstream tasks can help the model achieve better performance. ASTs and natural language both have a positive impact on the performance of the model, regardless of the downstream task.

\subsubsection*{RQ3: Is the ability of utilizing more unlabeled data an advantage of SPT-Code?}
\label{section:rq_pre_train_data}

As we know, CodeBERT, GraphCodeBERT and T5-learning are pre-trained on a subset of CodeSearchNet, whereas we can pre-train on the entire dataset. Therefore there is a question of whether the ability to utilize more data for pre-training also gives SPT-Code an unfair advantage. To ensure a fair comparison, we therefore pre-train SPT-Code only on the same data as CodeBERT and GraphCodeBERT (i.e., CSN$_{\textup{w/ doc}}$), and also on the same data as T5-learning (i.e., CSN$_{\textup{Java}}$). Notice, however, that here only the code from the labeled data is utilized and not the labels, even on the CSN$_{\textup{w/ doc}}$ where all samples have labels. So in this case, although we are using the data set of the same size, it actually utilize less information than they do. The results are displayed in the last group of Table~\ref{table:ablation_mixed}.

Comparing the results of SPT-Code with those of ``CSN$_{\textup{w/ doc}}$'' and ``CSN$_{\textup{Java}}$'' in Table~\ref{table:ablation_mixed}, we find that there is a significant performance decrease when the data used for pre-training is shrunk from 6.4M to 2.3M or 1.5M. However, considering the other pre-training models, it still maintains an advantage or is at a comparable level in performance. Therefore, we can conclude that SPT-Code is superior given the same amount of pre-trained data. In addition, the ability to use more unlabeled pre-training data can help SPT-Code achieve higher performance. This ensures that SPT-Code has a very advantageous scalability at the data level compared to the other pre-training models for source code. 

\subsubsection*{RQ4: How would the size of the training data for fine-tuning SPT-Code influence its performance on downstream tasks?}
\label{section:rq_fine_tune_data}

\begin{figure}[htbp]
  \centering
  \subfloat[Results on Java CodeSearchNet code summarization.]
    {\includegraphics[width=0.42\textwidth]{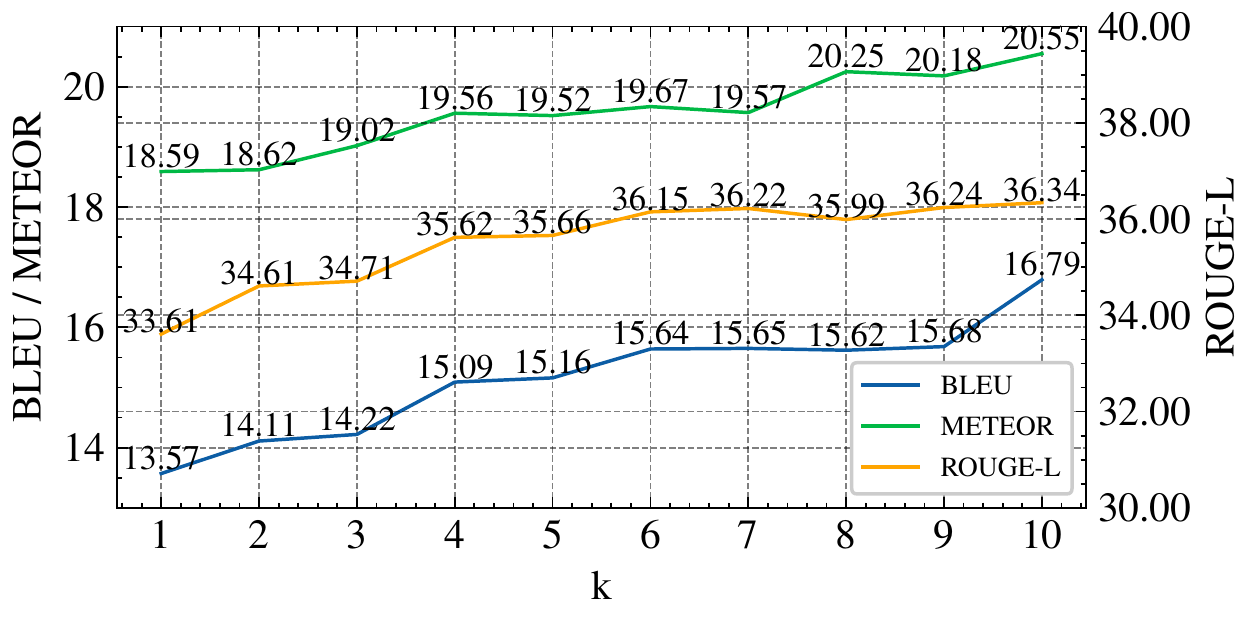}\label{figure:summarization_subset}}\\
  \subfloat[Results on BFP$_{\textup{medium}}$ bug fixing.]
    {\includegraphics[width=0.42\textwidth]{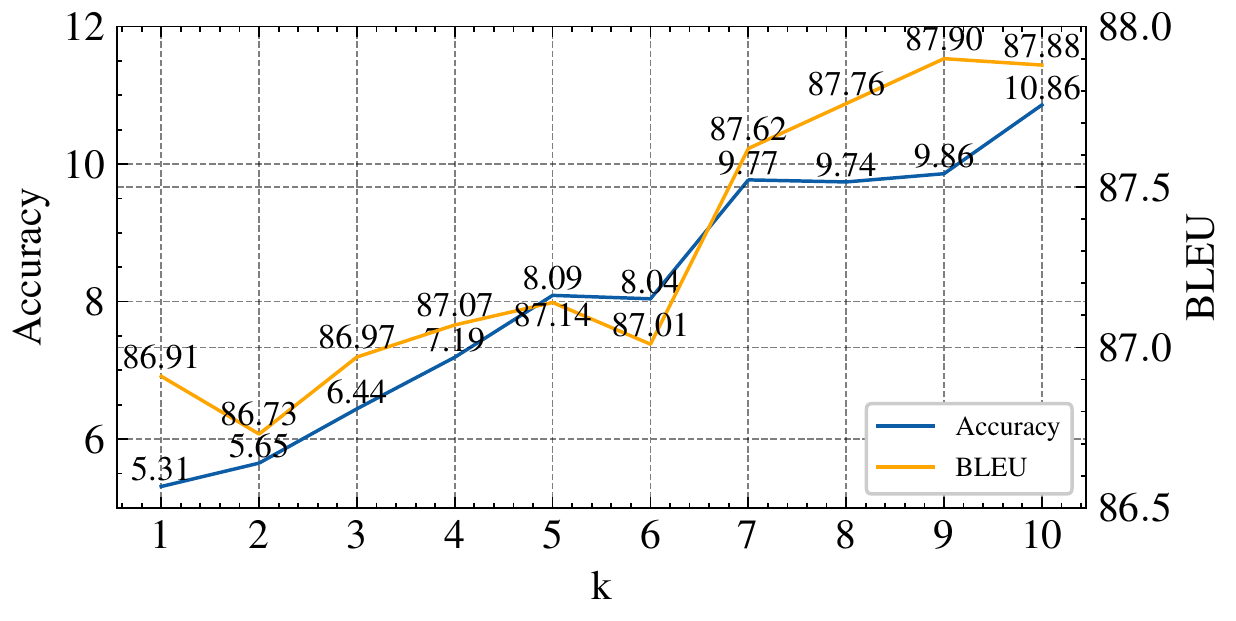}\label{figure:bug_fix_subset}}
  \caption{Results when $k$ goes from 1 to 10.}
  \Description{When the size of training set is different, results of the model.}
  \label{figure:data_ablation}
\end{figure}

To answer this question, we plot learning curves by varying the amount of (task-specific) training data used to fine-tune SPT-Code.
Owing to space limitations, we only report results of code summarization on Java in CodeSearchNet, and bug fixing on BFP$_{\textup{medium}}$. We select $k/10$ data from the training set each time for training, and then test on the same entire testing set. The results are shown in Figure~\ref{figure:data_ablation}.

It is easy to see that the performance shows a decreasing trend in all evaluation metrics as $k$ decreases (i.e., the size of the training set decreases). It proves that even for pre-trained models, the size of the training set plays a key role in the performance as well. Meanwhile, considering Table~\ref{table:results_summarization_csn}, we see that for code summarization, the results of SPT-Code are comparable to that of CodeBERT when the training set size is reduced to 1/10, and the performance is similar to or even higher than GraphCodeBERT when it is reduced to 2/10. Further, recalling Table~\ref{table:results_bug_fix}, we can see that when the training set of BFP$_{\textup{medium}}$ is reduced to 1/10 of the original size, SPT-Code's results are still higher than other pre-trained models, and when it is reduced to half of the original size, SPT-Code's performance can reach the performance of the best baseline model, i.e., S2S+COPYSPAN.

The conclusion is that although there is an inevitable drop in SPT-Code's performance as the size of the training data decreases during fine-tuning, the performance of SPT-Code is comparable to that of the other models when the data used for fine-tuning SPT-Code is reduced to a very small size. This also implies the robustness of SPT-Code.

\subsection{Quantitative Analysis}
\label{section:quantitative_analysis}

We collect the output of SPT-Code and baselines for a wide range of test samples on each downstream task. Then we invited five participants, all of whom are graduate students in software engineering who are themselves not authors of this paper. For each downstream task, the five students are randomly assigned to the same number of samples. If there are multiple datasets for that downstream task, the samples of multiple datasets are the same. For downstream tasks that use Acc as a metric, such as code completion, we focus on the similarity between incorrect and correct answers, and the extent to which the incorrect answer can be helpful when all models fail. The results are shown in Table~\ref{table:results_quantitative}. 
\begin{table}[htbp]
    \centering
    \caption{Results of quantitative analysis.}
    \label{table:results_quantitative}
    \footnotesize
    \begin{tabular}{lrrrr}
        \toprule
            Dowstream Tasks    & \# Sample & \# Better & \# Comparable & \# Worse \\
        \midrule
            Code Summarization & 800       & 328       & 368           & 104      \\
            Code Completion    & 600       & 312       & 233           & 55       \\
            Bug Fixing         & 200       & 89        & 75            & 36       \\
            Code Translation   & 200       & 83        & 65            & 52       \\
            Code Search        & 300       & 61        & 185           & 54       \\
        \bottomrule
    \end{tabular}
\end{table}

\subsection{Qualitative Analysis}
\label{section:qualitative_analysis}

By asking the participants in the previous subsection and browsing the output of each model by ourselves, we found that in comparison, SPT-Code can capture the semantic information of identifiers within code more accurately than other pre-trained models, and it can capture the semantic information of code segment globally instead of limiting to a certain region. The extracted code semantics are relatively more widely and evenly distributed in the code segment.

In the case of code summarization, for example, SPT-Code provides more complete and accurate descriptions of the method's overall functionality. Table~\ref{table:qualitative_example} shows summaries generated by different models for an example Java method in CodeSearchNet data. 
\begin{table}
    \centering
    \caption{Qualitative example of SPT-Code and baselines.}
    \footnotesize
    \begin{tabular}{l}
        \toprule
            \begin{lstlisting}
void emitLoop() {
    for ( ; ; ) {
        AppendOnlyLinkedArrayList<Object> q;
        q = queue;
        if (q == null) {
            emitting = false;
            return; }
        queue = null;
        q.forEachWhile(this); } }
            \end{lstlisting} \\
        \midrule
            {\color{teal} CodeBERT}: the queue is being destroyed \\
            {\color{cyan} GraphCodeBERT}: emits the next loop in the queue \\
            {\color{orange} T5-learning}: this method is called when the queue \\
        \hline
            {\color{purple} SPT-Code}: emits all of the elements in this queue \\
            {\color{purple} \ -w/o AST}: emit the linked list \\
            {\color{purple} \ -w/o NL}: emits all entries in the queue \\
        \hline
            {\color{red} Human Written}: loops until all notifications in the queue has been processed \\
        \bottomrule
    \end{tabular}
    \label{table:qualitative_example}
\end{table}
We find that the ``queue'' object is recognized by all of these models. CodeBERT and T5-learning fail to capture the operation conducted on the queue, i.e., ``emits/emit'', but GraphCodeBERT and SPT-Code does. Compared to SPT-Code, although GraphCodeBERT also understand the relationship between the operation and the object, it fails to choose appropriate words for describing the operation. The reason may be that the data flow can indeed partially (but not fully) capture code structure information. Another interesting observation that may support this inference is that all models considering structure information generate ``in the queue'' correctly, i.e., GraphCodeBERT, SPT-Code, and SPT-Code-w/o NL, and models using AST all generate accurate, readable and smooth summaries. It also implies that data flow is less effectiveness than AST. 

\section{Threats to Validity}
\label{section:threats}


\subsubsection*{Construct Validity} Like many existing code pre-training models, SPT-Code uses CodeSearchNet for pre-training. Since CodeSearchNet is also used in the evaluation of code summarization and code search, it is possible that the samples from the test sets for these two tasks have already been used for pre-training. This is not fair to methods that have not been pre-trained with CodeSearchNet, such as NeuralCodeSum in CodeSearchNet code summarization. We also recognize the impact on the results of not removing duplicates. However, our decision of not removing replicates is based on two considerations: (1) ensuring fairness of the comparison: the three code pre-training baselines were pre-trained with all data from CodeSearchNet without duplicate removal; (2) SPT-code is not affected by duplicate data: on one hand, downstream tasks that use pre-training dataset are CSN code summarization and code search, while the pre-training tasks designed for SPT-Code do not use the docstring, i.e., in testing, SPT-Code does not have the advantage of generating more accurate summaries because it has seen a piece of the tested code during pre-training, or has the advantage of better searching for code based on natural language. On the other hand, other downstream tasks that do not use CodeSearchNet have little overlap with CSN.

For rigorous consideration, we remove all test sets in CodeSearchNet and code duplicated with test sets of other downstream tasks\footnote{By using tools provided by Allamanis~\cite{allamanis2019adverse}, we find 9 in classical code summarization, 2 in $\textup{BFP}_\textup{medium}$ and code completion, respectively.}. Then re-pre-train and fine-tune the SPT-Code on three downstream tasks, i.e., JCSD, code completion and $\textup{BFP}_\textup{medium}$ bug fixing. It is found that the results of SPT-Code decrease very little after removing these duplicates\footnote{Results decrease 0.01 BLEU on JCSD, 0.03 Acc@1 on code completion and 0.07 Acc on $\textup{BFP}_\textup{medium}$}. Moreover, it is still not sure whether the change in results is due to the smaller pre-training dataset (shrunk by about 1/10) or the removal of duplicates.

\subsubsection*{Internal Validity} It is widely agreed that hyperparameters have a significant impact on the performance of deep learning models, but hyperparameters of SPT-Code are not tuned experimentally and are set empirically. Therefore, other hyperparameter settings may yield better results.

\subsubsection*{External Validity} Another threat posed by using CodeSearchNet as our pre-training dataset is that CodeSearchNet data is not balanced across six programming languages, which can be seen in Table~\ref{table:csn_statistics}, so our model may not perform the same on different languages, and we cannot guarantee the validity of SPT-Code for programming languages other than these six.

\section{Conclusion}
\label{section:conclusion}

We presented SPT-Code, a large model for source code based on an encoder-decoder architecture. 
First, we design three code-specific pre-training tasks to pre-train SPT-Code. Secondly, we propose a new input representation whose is the first method that take into account both natural language and AST form of code, where we also propose a improved version of the AST traversal method, X-SBT. Both our pre-training tasks and input representation allow SPT-Code to be pre-trained on a completely unlabeled dataset.
SPT-Code was then fine-tuned on five code-related downstream tasks. Results indicate that fine-tuning SPT-Code enables it to achieve the state-of-the-art performance on five code-related downstream tasks. Ablation experiments reveal that the three pre-training tasks have different degrees of impact on different downstream tasks, and AST and natural language input also helped improve SPT-Code's performance. To facilitate future research, we also make our code and other artifacts publicly available at \url{https://github.com/NougatCA/SPT-Code}.

\begin{acks}
This work is supported by National Natural Science Foundation of China (61802167, 61802095), Natural Science Foundation of Jiangsu Province, China (BK20201250), Cooperation Fund of Huawei-NJU Creative Laboratory for the Next Programming, and NSF award 2034508. We thank Alibaba Cloud for its high-efficient AI computing service from EFlops Cluster. We also thank the reviewers for their helpful comments. Chuanyi Li and Jidong Ge are the corresponding authors.
\end{acks}

\bibliographystyle{ACM-Reference-Format}
\bibliography{refs}










\end{document}